\documentclass[journal,twocolumn]{IEEEtran}
\usepackage{}
\usepackage{graphicx}
\usepackage{epstopdf}
\usepackage{amssymb}
\usepackage{amsfonts}
\usepackage{amsmath}
\usepackage{algorithm}
\usepackage{algorithmic}
\usepackage{subeqnarray}
\usepackage{cases}
\usepackage{bm}
\usepackage{color}
\usepackage{subfigure,amsmath,amssymb,cite}
\usepackage{float}
\ifCLASSINFOpdf
\else
\fi
\begin{document}

%\title{Coordinated Hybrid Precoding and Phase Shift for Intelligent Reconfigurable Surface-aided Directional Modulation Network}
\title{Performance Analysis of Wireless Network Aided by Discrete-Phase-Shifter IRS}
\author{Rongen Dong, Yin Teng, Zhongwen Sun, Jun Zou, Mengxing Huang,\\
 Jun Li, Feng Shu, and Jiangzhou Wang,\emph{ Fellow, IEEE}

%\thanks{Copyright (c) 2015 IEEE. Personal use of this material is permitted. However, permission to use this material for any other purposes must be obtained from the IEEE by sending a request to pubs-permissions@ieee.org.}
\thanks{This work was supported in part by the National Natural Science Foundation of China (Nos. 62071234, 62071289, and 61972093), the Hainan Major Projects (ZDKJ2021022), the Scientific Research Fund Project of Hainan University under Grant KYQD(ZR)-21008, and the National Key R\&D Program of China under Grant 2018YFB180110 \emph{(Corresponding authors: Feng Shu and Mengxing Huang)}.}
%%\thanks{Tong Shen,~Jin Wang,~and Feng Shu are with the School of Electronic and Optical Engineering, Nanjing University of Science and Technology, 210094, CHINA. (Email: shufeng0101 @163.com). }
\thanks{Rongen Dong, Zhongwen Sun, and Mengxing Huang are with the School of Information and Communication Engineering, Hainan University, Haikou, 570228, China.}
%\thanks{Baihua Shi is with the School of Electronic and Optical Engineering, Nanjing University of Science and Technology, Nanjing, 210094, China.}

\thanks{Yin Teng, Jun Zou, and Jun Li are with the School of Electronic and Optical Engineering, Nanjing University of Science and Technology, 210094, China.}
\thanks{Feng Shu is with the School of Information and Communication Engineering, Hainan University, Haikou, 570228, China, and also with the School of Electronic and Optical Engineering, Nanjing University of Science and Technology, Nanjing, 210094, China (e-mail: shufeng0101@163.com).}

\thanks{Jiangzhou Wang is with the School of Engineering, University of Kent, Canterbury CT2 7NT, U.K. (e-mail: {j.z.wang}@kent.ac.uk).
}

% <-this % stops a space
%
}

\maketitle

\begin{abstract}
Discrete phase shifters of intelligent reflecting surface (IRS)  generates phase quantization error (QE) and degrades the receive performance at the receiver. To make an analysis of the performance loss caused by IRS with phase QE, based on the law of large numbers, the closed-form expressions of signal-to-noise ratio (SNR) performance loss (PL), achievable rate (AR), and bit error rate (BER) are successively derived under line-of-sight (LoS) channels and Rayleigh channels. Moreover, based on the Taylor series expansion, the approximate simple closed form of PL of IRS with approximate QE is also given. The simulation results show that the performance losses of SNR and AR decrease as the number of quantization bits increase, while they gradually increase with the number of IRS phase shifter elements increase. Regardless of LoS channels or Rayleigh channels,
when the number of quantization bits is larger than or equal to 3, the performance losses of SNR and AR are less than 0.23dB and 0.08bits/s/Hz, respectively, and the BER performance degradation is trivial. In particular, the performance loss difference between IRS with QE and IRS with approximate QE is negligible when the number of quantization bits is not less than 2.

\end{abstract}
\section{Introduction}
With the rapid development of wireless networks, the demands for high rate, high quality, and ubiquitous wireless services will result in high energy consumption like the fifth generation (5G) systems \cite{Nguyen2022Achievable}. To achieve an innovative, energy-efficient and low-cost wireless network, intelligent reflecting surface (IRS) has emerged as a new and promising solution.
IRS,  consisting of a large number of low-cost passive reflective elements integrated on a plane, can significantly enhance the performance of wireless communication networks by intelligently reconfiguring the wireless propagation environment \cite{Wu2020Towards, Tang2020MIMO, Gao2021Reflection}. There are heavy research activities on the investigation of  various IRS-aided wireless networks \cite{Huang2019Reconfigurable, Guan2020Intelligent, Dong2022Beamforming, Hong2020Artificial, Wang2022Beamforming, Zhu2022Intelligent, Ahmed2022Reconfigurable, Mei2021Performance, Yang2020Secrecy}.

Assuming all channels are the line-of-sight (LoS) channels, the authors in \cite{Fang2021Joint}  maximized the secrecy rate (SR) by jointly optimizing IRS phases, and  the trajectory  and power control of unmanned aerial vehicle, based on the successive convex approximation, and the SR was significantly improved with the assistance of IRS.
In \cite{ShuEnhanced2021}, a secure IRS-aided directional modulation network was investigated, and two alternating iterative algorithms, general alternating iterative and null-space projection, were proposed to maximize the SR. An IRS-assisted downlink multi-user multi-antenna system in the absence of direct links between the base station (BS) and user was proposed in \cite{Di2020Hybrid}, a hybrid beamforming scheme with continuous digital beamforming for the BS and discrete analog beamforming for the IRS was proposed to maximize sum-rate. In \cite{Choi2021Multiple}, the phase shifters of multiple IRSs were optimized to maximize rate, based on the least-squares method, the substantial rate gains were achieved compared to the baseline schemes. The problem of joint active and passive beamforming optimization for an IRS-aided downlink multi-user multiple-input multiple-output (MIMO) system was investigated in \cite{Rehman2021Joint}, where a vector approximate message passing algorithm was proposed to optimize the IRS phase shifts.
In \cite{Han2022Double}, the transmit covariance matrix and passive beamforming matrices of the two cooperative IRSs were jointly optimized to maximize rate, and a novel low-complexity alternating optimization algorithm was presented.

Actually, there are many works focusing on the beamforming methods in the Rayleigh channels. An IRS-assisted multiple-input single-output (MISO) system without eavesdropper's channel state information (CSI) was proposed in\cite{Wang2020Intelligent}, in order to enhance the security, the oblique manifold method and minorization-maximization algorithms were proposed to jointly optimize the precoder and IRS phase shift. In \cite{Wu2019Beamforming}, the continuous transmit beamforming at the access point (AP) and discrete reflect beamforming at the IRS were jointly optimized to minimize the transmit power at AP. An efficient alternating optimization algorithm was proposed and near-optimal performance was achieved.
An IRS-aided secure multigroup multicast MISO communication system was proposed in \cite{Shi2021Secure}, and the semidefinite relaxation scheme and a low-complexity algorithm based on second-order cone programming were designed to minimize the transmit power.
In \cite{Zhang2020Sum}, based on the Arimoto-Blahut algorithm, the source precoders and IRS phase shift matrix in the full-duplex MIMO two-way communication system were optimized to maximize the sum rate.
A fast converging alternating algorithm to maximize the sum rate was proposed in \cite{Shen2020Beamforming}. Compared to the algorithm in \cite{Zhang2020Sum}, the proposed algorithm achieved a faster convergence rate and lower computational complexity.
In \cite{Pan2020Multicell}, the authors proposed to invoke an IRS at the cell boundary of multiple cells to assist the downlink transmission to cell-edge users, and the precoding matrices at the BSs and IRS phase shifts were jointly optimized to maximize the weighted sum rate of all users.

Similar to that discrete-quantized radio frequency phase shifter would cause performance loss in \cite{Li2019Performance, Wei2021Secure, Dong2022Performance}, using discrete-phase shifters in IRS will  also  result in performance loss in IRS-aided wireless network \cite{Di2020Hybrid, You2020Channel, Wu2019Beamforming}. Choosing a proper number of quantization bits for discrete-phase shifters with a given performance loss will  provide a reference for striking a good balance between hardware  circuit cost and performance loss. However, to the best of our knowledge, there is no research work on performance loss analysis  of the IRS with  discrete-phase shifters.
%to analyze the performance loss caused by finite-quantized RF phase shifter,  in \cite{Li2019Performance},  the authors derived the performance loss closed-form expression of the SINR and secrecy rate (SR) in analog beamforming structure. The performance loss analysis of hybrid DM with mixed phase shifters also have been analyzed in \cite{Dong2022Performance}.
Thus, in what follows, we will present an analysis on impact of discrete-phase shifters on the performance of IRS-aided wireless network system in this paper. Our main contributions are summarized as follows:
\begin{enumerate}

\item To make an analysis of performance loss caused by discrete-phase shifters, an IRS-aided wireless network is considered. We assume that all channels are LoS channels. Based on the law of large numbers, the closed-form expressions of signal-to-noise ratio (SNR) performance loss, achievable rate (AR), and bit error rate (BER) are successively derived. Simulation results show that the performance losses of SNR and AR gradually decrease as the number of quantization bits increase, while they gradually increase as the number of IRS phase shifter increases. When the number of quantization bits is equal to 3, the performance losses of SNR and AR are respectively less than 0.23dB and 0.08bits/s/Hz, and the BER performance degradation is negligible.

\item In the Rayleigh fading channels, with the weak law of large numbers and the Rayleigh distribution, the closed-form expressions of SNR  performance loss (PL) is  derived while AR and BER  with PL are given. In addition, based on the Taylor series expansion, the simple approximate performance loss  (APL) expression of SNR is derived whereas  AR and BER  with APL are given. Simulation results show that the SNR, AR and BER PL tendencies in the Rayleigh channels are similar to those in LoS channels.  That is, 3-bit phase shifters are sufficient to achieve an omitted performance loss.  %compared to the case of LoS channels, a lower SNR performance loss and worse AR and BER are achieved
    In particular, the approximate simple expression of performance loss makes a good approximation to  the true performance loss when the number of quantization bits is larger than or equal to 2.
\end{enumerate}

The remainder of this paper is organized as follows. Section II describes the system model of a typical IRS-aided three-node wireless network.
The performance loss derivations in the LoS  and Rayleigh channels are presented in Section III and Section IV, respectively. Numerical simulation results are presented in Section V. Finally, we draw conclusions in Section VI.

{\bf Notations:} throughout this paper, boldface lower case and upper case letters represent vectors and matrices, respectively. Signs $(\cdot)^T$, $(\cdot)^H$, $(\cdot)^{-1}$, $\|\cdot\|_2$, and $|\cdot|_m$ denote the transpose operation, conjugate transpose operation, inverse operation, 2-norm operation, and $m$-th element absolute value operation, respectively. The symbol $\mathbb{C}^{N\times N}$ denotes the space of $N\times N$ complex-valued matrix. The notation $\textbf{I}_N$ represents the $N\times N$ identity matrix. %$P(\cdot)$ represents the PDF.

\section{system model}
\begin{figure}[htbp]
\centering
\includegraphics[width=0.45\textwidth]{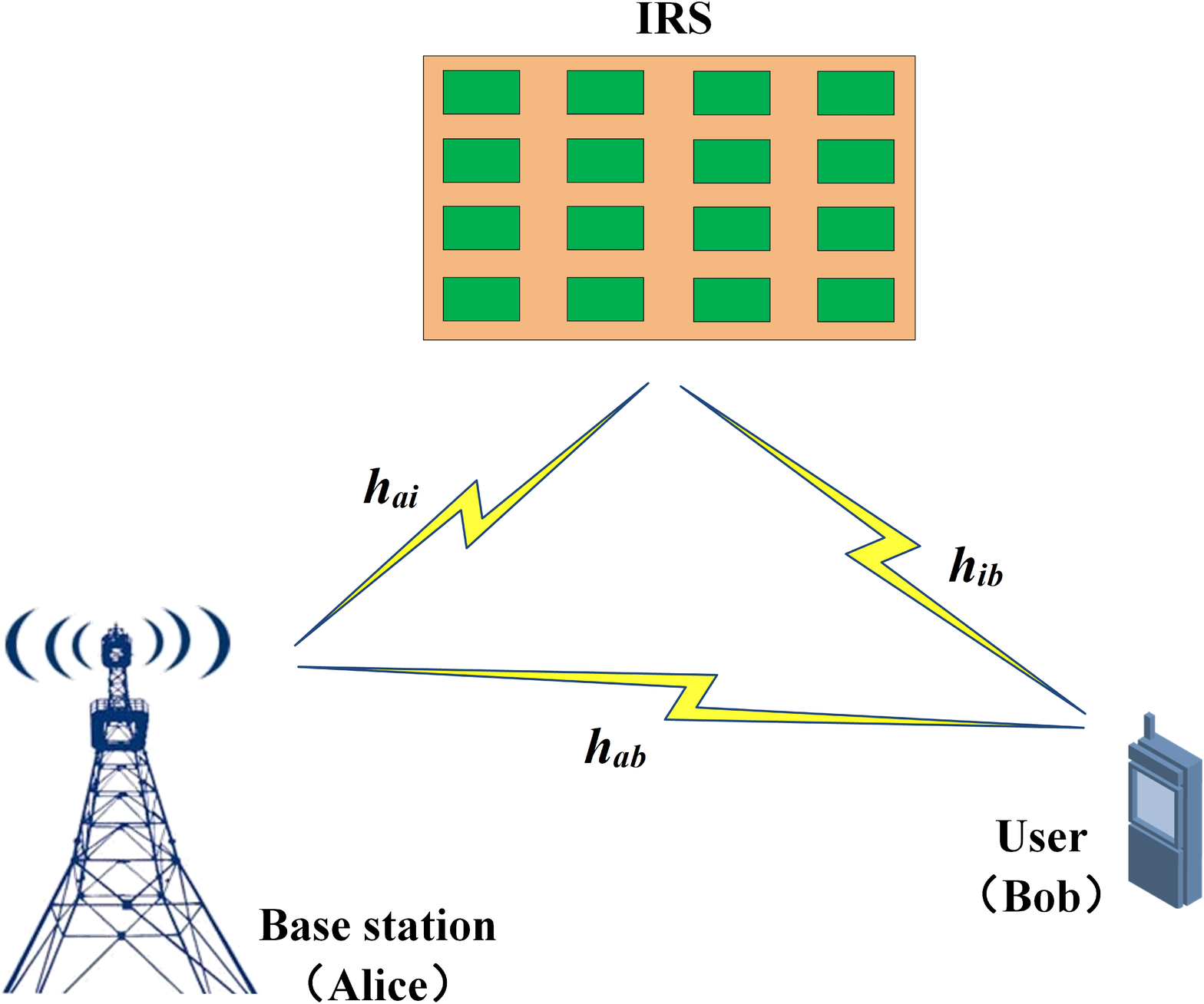}\\
\caption{System model of IRS-aided wireless network.}\label{model}
\end{figure}

%\subsection{IRS-aided directional modulation system model}
As shown in Fig.~\ref{model}, an IRS-aided wireless network system is considered. Herein, the base station (Alice) and user (Bob) are equipped with single antenna.  The IRS is equipped with $M$ low-cost passive reflecting elements and reflects signal only one time slot. The Alice$\rightarrow$IRS, Alice$\rightarrow$Bob, and IRS$\rightarrow$Bob channels are the LoS or Rayleigh channels.

% bmeps -c model.jpg model.eps(convert a to b)

The transmit signal at Alice is given by
\begin{align}
s=\sqrt{P_a}x,
\end{align}  %N为总天线数
where $P_a$ denotes the total transmit power, $x$ is the confidential message and satisfies $\mathbb{E}[\|x\|^2]=1$.

Taking the path loss into consideration, the received signal at Bob is
\begin{align}\label{y_b}
y_b&=
\left(\sqrt{g_{aib}}\textbf{h}^H_{ib}\boldsymbol{\Theta}\textbf{h}_{ai}+\sqrt{g_{ab}}\textbf{h}^H_{ab}\right)s+n_b\nonumber\\
&=\left(\sqrt{g_{aib}P_a}\textbf{h}^H_{ib}\boldsymbol{\Theta}\textbf{h}_{ai}+\sqrt{g_{ab}P_a}h^H_{ab}\right)x+n_b,
\end{align}
where $g_{aib}=g_{ai}g_{ib}$ represents the equivalent path loss coefficient of Alice$\rightarrow$IRS channel and IRS$\rightarrow$Bob channel, and $g_{ab}$ is the path loss coefficient of Alice$\rightarrow$Bob channel. $n_b$ denotes the additive while Gaussian noise (AWGN) at Bob with the distribution $\mathcal {C}\mathcal {N}\sim(0, \sigma^2)$.
$\boldsymbol{\Theta}=\text{diag}\left(e^{j\phi_1}, \cdots, e^{j\phi_m}, \cdots, e^{j\phi_M}\right)$ represents the diagonal reflection coefficient matrix of IRS, where $\phi_m\in [0,2\pi)$ denotes the phase shift of reflection element $m$.
$\textbf{h}_{ai}\in\mathbb{C}^{M\times 1}$, $\textbf{h}_{ab}=h_{ab}\in\mathbb{C}^{1\times 1}$, and $\textbf{h}_{ib}\in\mathbb{C}^{M\times 1}$ are the Alice$\rightarrow$IRS, Alice$\rightarrow$Bob, and IRS$\rightarrow$Bob channels, respectively.

\section{Performance Loss Derivation and Analysis in the LoS Channels}
In this section, it is assumed that all channels are the LoS channels. The use of IRS with discrete phase shifters may lead to phase quantization errors. In what follows, we  will make a comprehensive investigation of the impact of IRS with discrete phase shifters on SNR, AR, and BER.

Defining $\textbf{h}_{ai}=\textbf{h}(\theta_{ai})$, $\textbf{h}_{ib}=\textbf{h}(\theta_{ib})$,
the steering vector  arrival or departure from IRS is
\begin{align}\label{h_theta}
\textbf{h}(\theta)=\left[e^{j2\pi\Psi_{\theta}(1)}, \dots, e^{j2\pi\Psi_{\theta}(m)}, \dots, e^{j2\pi\Psi_{\theta}(M)}\right]^T,
\end{align}
and the phase function $\Psi_{\theta}(m)$ is given by
\begin{align}
\Psi_{\theta}(m)\buildrel \Delta \over =-\frac{(m-(M+1)/2)d \cos\theta}{\lambda}, m=1, \dots, M,
\end{align}
where $m$ denotes the $m$-th antenna, $d$ is the  spacing of adjacent transmitting antennas, $\theta$ represents the direction angle of arrival or departure, and $\lambda$ represents the wavelength.

The receive signal (\ref{y_b})  can be casted as
\begin{align}\label{y_b11}
y^{LoS}_b&=\left(\sqrt{g_{aib}P_a}\textbf{h}^H(\theta_{ib})\boldsymbol{\Theta}\textbf{h}(\theta_{ai})+
\sqrt{g_{ab}P_a}h^H_{ab}\right)x+n_b \nonumber\\
&=\sqrt{g_{aib}P_a}\left({\sum\limits_{m = 1}^M e^{j\left(-2\pi \Psi_{\theta_{ib}}(m)+
\phi_m+2\pi \Psi_{\theta_{ai}}(m)\right)} }\right)x\nonumber\\
&~~+\sqrt{g_{ab}P_a}h^H_{ab}x+n_b.
\end{align}
If the phase shifter at IRS is continuous, and the transmit signal at Alice is forwarded perfectly to Bob by the IRS, the $m$-th phase shift at IRS can be designed as follows
\begin{align}
\phi_m=2\pi \Psi_{\theta_{ib}}(m)-2\pi \Psi_{\theta_{ai}}(m),
\end{align}
then (\ref{y_b11}) can be converted to
\begin{align}\label{y_b_NQ}
y^{LoS}_b=\sqrt{g_{aib}P_a}Mx+\sqrt{g_{ab}P_a}h^H_{ab}x+n_b.
\end{align}

\subsection{Derivation of Performance Loss in LoS Channels}
Assuming the discrete phase shifters is employed by IRS, and the discrete phases per phase shifters at IRS employs a $k$-bit phase quantizer, each reflection element's phase feasible set is
\begin{align}
\Omega=\left\{\frac{\pi}{2^k}, \frac{3\pi}{2^k},  \cdots, \frac{(2^{k+1}-1)\pi}{2^k}\right\}.
\end{align}
Assuming that $ \phi_m$ is the desired continuous phase of the $m$-th element at IRS, and the final discrete phase is chosen from phase feasible set $\Omega$, which is given by
\begin{align}
\overline{\phi_m}=\mathop{\arg\min}\limits_{{\overline{\phi_m}\in \Omega}}  \|\phi_m-\overline{\phi_m}\|_2.
\end{align}
In general, $\overline{\phi_m}\neq \phi_m$, which means that phase mismatching may lead to performance loss at Bob. Let us define the $m$-th phase quantization error at IRS as follows
\begin{align}
\Delta \phi_m=\phi_m-\overline{\phi_m}.
\end{align}
It is assumed that the above phase quantization error follows uniform distribution with probability density function (PDF) as follows
\begin{equation}\label{f0}
  f(x)=\left\{
             \begin{array}{ll}
             \frac{1}{2\Delta x}, & \hbox{~$x\in[-\Delta x, \Delta x]$,}\\
             0, & \hbox{~otherwise},
             \end{array}
           \right.
\end{equation}
where
\begin{align}
\Delta x=\frac{\pi}{2^k}.
\end{align}

In the presence of phase quantization error, the receive signal (\ref{y_b}) becomes
\begin{align}\label{y_b-error}
\widehat{y_b}^{LoS}
&=\left(\sqrt{g_{aib}P_a}\textbf{h}^H(\theta_{ib})\boldsymbol{\Theta}\textbf{h}(\theta_{ai})+
\sqrt{g_{ab}P_a}h^H_{ab}\right)x+n_b \nonumber\\
&=\sqrt{g_{aib}P_a}\left({\sum\limits_{m = 1}^M e^{j\left(-2\pi \Psi_{\theta_{ib}}(m)+
\overline{\phi_m}+2\pi \Psi_{\theta_{ai}}(m)\right)} }\right)x\nonumber\\
&~~+\sqrt{g_{ab}P_a}h^H_{ab}x+n_b\nonumber\\
&=\sqrt{g_{aib}P_a}\left({\sum\limits_{m = 1}^M e^{j\bigtriangleup \phi_m} }\right)x+
\sqrt{g_{ab}P_a}h^H_{ab}x+n_b.
\end{align}
Observing the above equation, it is apparently that if and only if $\bigtriangleup \phi_m=0$, the phase alignment at user is achieved to realize the optimal coherent combining gain $M^2$. Due to  the use of finite phase shifting, in general,  $\bigtriangleup \phi_m$ is random and is unequal to zero, this means that the combining gain is lower than or far less than $M^2$. In other words, the receive performance decays.

In accordance with the law of large numbers in \cite{Wasserman2004All} and (\ref{f0}), we can obtain
\begin{align}\label{M1}
\frac{1}{M}{\sum\limits_{m = 1}^M e^{j\Delta \phi_m}}&\approx \mathbb{E}\left(e^{j\Delta \phi_m}\right)\nonumber\\
&=\int_{-\Delta x}^{\Delta x}e^{j\Delta \phi_m}f\left(\Delta \phi_m\right)d\left(\Delta \phi_m\right) \nonumber\\
&=\int_{-\Delta x}^{\Delta x}\frac{e^{j\Delta \phi_m}}{2\Delta x}d\left(\Delta \phi_m\right) \nonumber\\
&=\frac{1}{2\Delta x}\int_{-\Delta x}^{\Delta x}\cos\left(\Delta \phi_m\right)d\left(\Delta \phi_m\right).
\end{align}
A further simplification of (\ref{M1}) yields
\begin{align}\label{delta}
\frac{1}{M}{\sum\limits_{m = 1}^M e^{j\Delta \phi_m}}&\approx \frac{\text{sin}\Delta_x}{\Delta_x}
=\text{sinc}\left(\frac{\pi}{2^k}\right).
\end{align}
Plugging (\ref{delta}) in (\ref{y_b-error}) yields
\begin{align}\label{y_b_QE}
\widehat{y_b}^{LoS}\approx \sqrt{g_{aib}P_a}M\text{sinc}\left(\frac{\pi}{2^k}\right)x+\sqrt{g_{ab}P_a}h^H_{ab}x+n_b.
\end{align}

In what follows, to simplify (\ref{y_b_QE}), we consider that the number of quantization bits is large, that is, $\bigtriangleup \phi_m $ goes to zero. Using the Taylor series expansion \cite{Moon1999Mathematical}, we have the following approximation
\begin{align}\label{Taylor}
\cos\left(\bigtriangleup \phi_m\right)\thickapprox 1-\frac{\bigtriangleup \phi^2_m}{2},
\end{align}
then (\ref{M1}) can be rewritten as
\begin{align}
\frac{1}{M}{\sum\limits_{m = 1}^M e^{j\Delta \phi_m}}&\approx
\frac{1}{2\Delta x}\int_{-\Delta x}^{\Delta x}\cos\left(\Delta \phi_m\right)d\left(\Delta \phi_m\right)\nonumber\\
&\approx \frac{1}{2\Delta x}\int_{-\Delta x}^{\Delta x}1-\frac{\bigtriangleup \phi^2_m}{2}d\left(\Delta \phi_m\right)\nonumber\\
&=\frac{1}{2\Delta x}\left(2\Delta x-\frac{1}{3}(\Delta x)^3\right)\nonumber\\
&=1-\frac{1}{6}\left(\frac{\pi}{2^k}\right)^2.
\end{align}
At this point, the receive signal at Bob under the approximate phase quantization error is
\begin{align}\label{y_b_AQE1}
\widetilde{y_b}^{LoS}\approx \sqrt{g_{aib}P_a}\Big(1-\frac{1}{6}\left(\frac{\pi}{2^k}\right)^2\Big)Mx+
\sqrt{g_{ab}P_a}h^H_{ab}x+n_b.
\end{align}

\subsection{Performance Loss  of SNR at Bob}
In accordance with (\ref{y_b_NQ}), the SNR expression with no PL, i.e., $k\rightarrow\infty$, is given by
\begin{align}
\text{SNR}^{LoS}=\frac{\left(\sqrt{g_{aib}P_a}M+\sqrt{g_{ab}P_a}h^H_{ab}\right)^2}{\sigma^2}.
\end{align}
From (\ref{y_b_QE}) and (\ref{y_b_AQE1}), the SNR PL and approximate PL (APL) are
\begin{align}
\widehat{\text{SNR}}^{LoS}=\frac{\left(\sqrt{g_{aib}P_a}M\text{sinc}\left(\frac{\pi}{2^k}\right)+
\sqrt{g_{ab}P_a}h^H_{ab}\right)^2}{\sigma^2},
\end{align}
and
\begin{align}
\widetilde{\text{SNR}}^{LoS}=\frac{\left(\sqrt{g_{aib}P_a}\left(1-\frac{1}{6}\left(\frac{\pi}{2^k}\right)^2\right)M+
\sqrt{g_{ab}P_a}h^H_{ab}\right)^2}{\sigma^2},
\end{align}
respectively, where $k$ is a finite positive integer.

Then the SNR PL and APL are given by
\begin{align}
\widehat{L}^{LoS}=\frac{\text{SNR}^{LoS}}{\widehat{\text{SNR}}^{LoS}}=
\frac{\left(\sqrt{g_{aib}}M+\sqrt{g_{ab}}h^H_{ab}\right)^2}
{\left(\sqrt{g_{aib}}M\text{sinc}\left(\frac{\pi}{2^k}\right)+\sqrt{g_{ab}}h^H_{ab}\right)^2},
\end{align}
and
\begin{align}
\widetilde{L}^{LoS}=\frac{\text{SNR}^{LoS}}{\widetilde{\text{SNR}}^{LoS}}=
\frac{\left(\sqrt{g_{aib}}M+\sqrt{g_{ab}}h^H_{ab}\right)^2}
{\left(\sqrt{g_{aib}}\left(1-\frac{1}{6}\left(\frac{\pi}{2^k}\right)^2\right)M+\sqrt{g_{ab}}h^H_{ab}\right)^2},
\end{align}
respectively.

\subsection{Performance Loss of Achievable Rate at Bob}
According to (\ref{y_b_NQ}), (\ref{y_b_QE}), and (\ref{y_b_AQE1}), the achievable rate at Bob with no PL, PL, and APL are given by
\begin{align}
R^{LoS}=\text{log}_2\left(1+\frac{\left(\sqrt{g_{aib}P_a}M+\sqrt{g_{ab}P_a}h^H_{ab}\right)^2}{\sigma^2}\right),
\end{align}
\begin{align}
\widehat{R}^{LoS}=\text{log}_2\left(1+\frac{\left(\sqrt{g_{aib}P_a}M\text{sinc}\left(\frac{\pi}{2^k}\right)+
\sqrt{g_{ab}P_a}h^H_{ab}\right)^2}{\sigma^2}\right),
\end{align}
and
\begin{align}
&\widetilde{R}^{LoS}=\nonumber\\
&\text{log}_2\left(1+\frac{\left(\sqrt{g_{aib}P_a}\left(1-\frac{1}{6}\left(\frac{\pi}{2^k}\right)^2\right)M+
\sqrt{g_{ab}P_a}h^H_{ab}\right)^2}{\sigma^2}\right),
\end{align}
respectively.
%The achievable rate loss is
%\begin{align}
%L_{R_{LoS}}&=\frac{R_{LoS}}{\widehat{R_{LoS}}}=\text{log}_2\left(\frac{Q}{\sigma^2}\right)
%\end{align}
%where
%\begin{align}
%Q&=\left(\sqrt{g_{aib}P_a}M+\sqrt{g_{ab}P_a}h_0\right)^2-\nonumber\\
%&~~~\left(\sqrt{g_{aib}P_a}M\text{sinc}\left(\frac{\pi}{2^k}\right)+\sqrt{g_{ab}P_a}h_0\right)^2\nonumber\\
%&=g_{aib}P_aM^2\left(1-\text{sinc}^2\left(\frac{\pi}{2^k}\right)\right)+
%2\sqrt{g_{aib}g_{ab}}P_aMh_0\left(1-\text{sinc}\left(\frac{\pi}{2^k}\right)\right)
%\end{align}

%The optimization problem can be formulated as follows
%\begin{subequations}
%\begin{align}\label{p0}
%&\max \limits_{\textbf{v}_{a},\bm{\Theta}}
%~~R_b\\
%&~~~\text{s.t.} ~~\|\textbf{v}_a\|^2=1, |\theta_i|=1, i=1,\cdots,M.
%\end{align}
%\end{subequations}
\subsection{Performance Loss  of BER at Bob}
In accordance with  \cite{Wasserman2004All}, the expression of BER is
\begin{align}
\text{BER}(z)\approx \beta Q\left(\sqrt{\mu z}\right),
\end{align}
where $\beta$ and $\mu$ depend on the type of approximation and the modulation type, $\beta$ represents the number of nearest neighbors to a constellation at the minimum distance, and $\mu$ is a constant that is related to minimum distance to average symbol energy, $z$ denotes the SNR per bit, $Q(z)$ represents the probability that a Gaussian random variable $x$ with mean zero and variance one exceeds the value $z$, i.e.,
\begin{align}
Q(z)=\int_{z}^{+\infty}\frac{1}{\sqrt{2\pi}}e^{\frac{-x^2}{2}}dx.
\end{align}

Assuming the quadrature phase shift keying (QPSK) is employed as the modulation scheme, according to (\ref{y_b_NQ}), (\ref{y_b_QE}), and (\ref{y_b_AQE1}), the BERs at Bob with no PL, PL, and APL are given by
\begin{align}
\text{BER}^{LoS}\approx Q\left(\sqrt{\frac{2\left(\sqrt{g_{aib}P_a}M+\sqrt{g_{ab}P_a}h^H_{ab}\right)^2}{\sigma^2}}\right),
\end{align}
\begin{align}
&\widehat{\text{BER}}^{LoS}\approx Q\left(\sqrt{\frac{2\left(\sqrt{g_{aib}P_a}M\text{sinc}\left(\frac{\pi}{2^k}\right)+
\sqrt{g_{ab}P_a}h^H_{ab}\right)^2}{\sigma^2}}\right),
\end{align}
and
\begin{align}
&\widetilde{\text{BER}}^{LoS}\approx \nonumber\\ &Q\left(\sqrt{\frac{2\left(\sqrt{g_{aib}P_a}\left(1-\frac{1}{6}\left(\frac{\pi}{2^k}\right)^2\right)M+
\sqrt{g_{ab}P_a}h^H_{ab}\right)^2}{\sigma^2}}\right),
\end{align}
respectively. This completes the derivations of the corresponding SNR performance loss, ARs and BERs with PL and APL in LoS channels.
\section{Performance Loss Derivation and Analysis in the Rayleigh Channels}
In this section, we make an analysis of the impact of  discrete phase shift of IRS on SNR, AR, and BER. The corresponding  SNR, AR, and BER performance loss expressions are derived in the Rayleigh fading channels.

\subsection{Derivation of Performance Loss in the Rayleigh Channels}
Assuming all channels are Rayleigh channels obeying the Rayleigh distribution, the corresponding PDF  is as follows
\begin{equation}\label{f}
  f_{\alpha}(x)=\left\{
             \begin{array}{ll}
             \frac{x}{\alpha^2}e^{-\frac{x^2}{2\alpha^2}}, & \hbox{~$x\in[0, +\infty)$,}\\
             0, & \hbox{~otherwise},
             \end{array}
           \right.
\end{equation}
where $\alpha>0$ represents the Rayleigh distribution parameter.

Assuming discrete phase shifters is employed by IRS, there is a phase quantization error due to the effect of phase mismatching, i.e., $\bigtriangleup \phi_m\neq0$, then the performance loss is incurred. Due to the phase mismatching of discrete phase shifters in IRS, the receive signal (\ref{y_b}) can be rewritten as
\begin{align}\label{y_b2}
\widehat{y}_b^{RL}&=\left(\sqrt{g_{aib}P_a}\textbf{h}^H_{ib}\boldsymbol{\Theta}\textbf{h}_{ai}+
\sqrt{g_{ab}P_a}h^H_{ab}\right)x+n_b\nonumber\\
&=\Big(\sqrt{g_{aib}P_a}{\sum\limits_{m = 1}^M
(|\textbf{h}^H_{ib}|_m|\textbf{h}_{ai}|_me^{j\bigtriangleup \phi_m}})+
\sqrt{g_{ab}P_a}|h^H_{ab}|\Big)x \nonumber\\
&~~~+n_b \nonumber\\
&=\Big(\sqrt{g_{aib}P_a}\Big(M\cdot\underbrace{\frac{1}{M}\sum\limits_{m = 1}^M
(\left|\textbf{h}^H_{ib}\right|_m\left|\textbf{h}_{ai}\right|_m\cos\left(\bigtriangleup \phi_m \right))}_{W}+ \nonumber\\
&~~~jM\cdot\underbrace{\frac{1}{M}{\sum\limits_{m = 1}^M (|\textbf{h}^H_{ib}|_m|\textbf{h}_{ai}|_m\sin(\bigtriangleup \phi_m))}}_{G}\Big)+\nonumber\\
&~~~\sqrt{g_{ab}P_a}\mathbb{E}\left(|h^H_{ab}|\right)\Big)x+n_b.
\end{align}

Using the weak law of large numbers, and the fact that all elements of $\textbf{h}_{ai}$  and $\textbf{h}_{ib}$ are independently identically distributed  Rayleigh distributions with parameters $\alpha_{ai}$ and  $\alpha_{ib}$, respectively,  and  their elements are independent of each other, we have
\begin{align}\label{G}
G&\approx \mathbb{E}\left(\left|\textbf{h}^H_{ib}\right|_m\left|\textbf{h}_{ai}\right|_m\sin(\bigtriangleup \phi_m)\right)
\nonumber\\
&=\iiint|h^H_{ib}|_m |h_{ai}|_m\sin(\bigtriangleup \phi_m)f_{\alpha_{ib}}(|h^H_{ib}|_m)f_{\alpha_{ai}}\left(|h_{ai}|_m\right) \nonumber\\
&~~\bullet f_1\left(\sin(\bigtriangleup \phi_m)\right)d\left(\bigtriangleup \phi_m\right)
d\left(|h_{ai}|_m\right)d\left(|h^H_{ib}|_m\right).
\end{align}
Since $|h^H_{ib}|$, $|h_{ai}|$, and $\bigtriangleup \phi_m $ are independent of each other, (\ref{G}) can be further converted to
\begin{align}\label{G1}
G&\approx 
\int_{0}^{+\infty}|h^H_{ib}|_mf_{\alpha_{ib}}(|h^H_{ib}|_m)\int_{0}^{+\infty}|h_{ai}|_mf_{\alpha_{ai}}\left(|h_{ai}|_m\right)\bullet\nonumber\\
&\int_{-\Delta x}^{\Delta x}\sin\left(\bigtriangleup \phi_m\right)
f\left(\bigtriangleup \phi_m)\right)
d\left(\Delta \phi_m\right)d\left(|h_{ai}|_m\right)d\left(|h^H_{ib}|_m\right)\nonumber\\
&=0.
\end{align}

Due to the fact that $|h^H_{ib}|$, $|h_{ai}|$, and $\bigtriangleup \phi_m $ are also independent of each other, similar to the derivation of (\ref{G}) and (\ref{G1}), we have
%\begin{figure*}[htbp]%ht放在顶部，hb放在底部
\begin{align}{\label{W1}}
W
&=\frac{1}{M}{\sum\limits_{m = 1}^M
\left(\left|\textbf{h}^H_{ib}\right|_m\left|\textbf{h}_{ai}\right|_m\cos(\bigtriangleup \phi_m)\right)}\nonumber\\
&\thickapprox
\mathbb{E}\left(\left|\textbf{h}^H_{ib}\right|_m\left|\textbf{h}_{ai}\right|_m\cos\left(\bigtriangleup \phi_m\right)\right)\nonumber\\
&=
\int_{0}^{+\infty}|h^H_{ib}|_mf_{\alpha_{ib}}\left(|h^H_{ib}|_m \right)\int_{0}^{+\infty}
|h_{ai}|_mf_{\alpha_{ai}}\left(|h_{ai}|_m \right)\bullet \nonumber\\
&~\int_{-\Delta x}^{\Delta x}\cos\left(\bigtriangleup \phi_m\right)
f\left(\bigtriangleup \phi_m\right)
d\left(\Delta \phi_m\right)d\left(|h_{ai}|_m\right)d\left(|h^H_{ib}|_m\right)\nonumber\\
&=\text{sinc}\left(\frac{\pi}{2^k}\right)\int_{0}^{+\infty}|h^H_{ib}|_mf_{\alpha_{ib}}\left(|h^H_{ib}|_m\right)
\int_{0}^{+\infty}|h_{ai}|_m \bullet \nonumber\\
&~~f_{\alpha_{ai}}\left(|h_{ai}|_m\right)d\left(|h_{ai}|_m\right)d\left(|h^H_{ib}|_m\right)\nonumber\\
&=\text{sinc}\left(\frac{\pi}{2^k}\right)\frac{\pi}{2}\alpha_{ai}\alpha_{ib}.
\end{align}

%\end{figure*}
Plugging (\ref{G1}) and (\ref{W1}) into (\ref{y_b2}) yields
\begin{align}\label{y_b_LQE1}
\widehat{y}_b^{RL}\approx &\left(\sqrt{g_{aib}P_a}M
\text{sinc}\left(\frac{\pi}{2^k}\right)\frac{\pi}{2}\alpha_{ai}\alpha_{ib}+
\sqrt{\frac{g_{ab}P_a\pi}{2}}\alpha_{ab}\right)x \nonumber\\
& +n_b,
\end{align}
where $\alpha_{ab}$ is the Rayleigh distribution parameter of channel from Alice to Bob.

To simplify (\ref{y_b_LQE1}), in terms of (\ref{Taylor}) and the fact that $|h^H_{ib}|$, $|h_{ai}|$, and $\bigtriangleup \phi_m^2 $ are independent of each other, we can obtain
%begin{figure*}[htbp]
\begin{align}{\label{W_Taylor}}
W&\thickapprox \frac{1}{M}{\sum\limits_{m = 1}^M
\left|\textbf{h}^H_{ib}\right|_m\left|\textbf{h}_{ai}\right|_m
\left(1-\frac{\bigtriangleup \phi^2_m}{2}\right)}\nonumber\\
&=\frac{1}{M}\sum\limits_{m = 1}^M\left|\textbf{h}^H_{ib}\right|_m\left|\textbf{h}_{ai}\right|_m -\frac{1}{M}
\sum\limits_{m = 1}^M\left|\textbf{h}^H_{ib}\right|_m\left|\textbf{h}_{ai}\right|_m \frac{\bigtriangleup \phi^2_m}{2}
\nonumber\\
&=\mathbb{E}\left(\left|\textbf{h}^H_{ib}\right|_m\left|\textbf{h}_{ai}\right|_m\right)-
\mathbb{E}\left(\left|\textbf{h}^H_{ib}\right|_m\left|\textbf{h}_{ai}\right|_m \frac{\bigtriangleup \phi^2_m}{2}\right)
\nonumber\\
&=\int_{0}^{+\infty}|h^H_{ib}|_mf_{\alpha_{ib}}\left(|h^H_{ib}|_m\right)
\int_{0}^{+\infty}|h_{ai}|_mf_{\alpha_{ai}}(|h_{ai}|_m)\bullet \nonumber\\
&~~~d\left(|h_{ai}|_m\right)d\left(|h^H_{ib}|_m\right)-\int_{0}^{+\infty}
|h^H_{ib}|_mf_{\alpha_{ib}}\left(|h^H_{ib}|_m\right)\bullet  \nonumber\\
&~~\int_{0}^{+\infty}|h_{ai}|_mf_{\alpha_{ai}}\left(|h_{ai}|_m\right)\int_{-\Delta x}^{\Delta x}\frac{\bigtriangleup \phi^2_m}{2}
f\left(\bigtriangleup \phi_m)\right)\bullet \nonumber\\
&~~~d\left(\Delta \phi_m\right)d\left(|h_{ai}|_m\right)d\left(|h^H_{ib}|_m\right)\nonumber\\
&=\frac{\pi}{2}\alpha_{ai}\alpha_{ib}-
\frac{1}{6}\left(\frac{\pi}{2^k}\right)^2\frac{\pi}{2}\alpha_{ai}\alpha_{ib}\nonumber\\
&=\left(1-\frac{1}{6}\left(\frac{\pi}{2^k}\right)^2\right)\frac{\pi}{2}\alpha_{ai}\alpha_{ib}.
\end{align}
%\hrulefill
%\end{figure*}
Plugging (\ref{W_Taylor}) into (\ref{y_b2}) yields
\begin{align}\label{y_b_LQE2}
\widetilde{y}_b^{RL}\approx &\Bigg(\sqrt{g_{aib}P_a}M
\Big(1-\frac{1}{6}\big(\frac{\pi}{2^k}\big)^2\Big)\frac{\pi}{2}\alpha_{ai}\alpha_{ib}+\nonumber\\
&\sqrt{\frac{g_{ab}P_a\pi}{2}}\alpha_{ab}\Bigg)x+n_b.
\end{align}

Assuming there is no quantization error, i.e., $\bigtriangleup \phi_m=0$, the receive signal (\ref{y_b_LQE1}) degrades to
\begin{align}\label{y_b_LNQ}
&y_b^{RL}
%=\Big(\sqrt{g_{aib}P_a}{\sum\limits_{m = 1}^M \left|\textbf{h}^H_{ib}\right|_m\left|\textbf{h}_{ai}\right|_m}+
%\sqrt{g_{ab}P_a}|h^H_{ab}|\Big)x+n_b \nonumber\\
%&\approx \big(\sqrt{g_{aib}P_a}\sum\limits_{m = 1}^M
%\mathbb{E}(|\textbf{h}^H_{ib}|_m|\textbf{h}_{ai}|_m)+
%\sqrt{g_{ab}P_a}\mathbb{E}(|h^H_{ab}|)\big)x +n_b \nonumber\\
%&=\Big(\sqrt{g_{aib}P_a}\sum\limits_{m = 1}^M\iint |h^H_{ib}|_m|h_{ai}|_mP(|h^H_{ib}|_m)P(|h_{ai}|_m)\bullet
%\nonumber\\
%&d(|h_{ai}|_m)d(|h^H_{ib}|_m)+\sqrt{g_{ab}P_a}\int |h^H_{ab}|P(|h^H_{ab}|)d{|h^H_{ab}|}\Big)x+n_b \nonumber\\
%&=\Big(\sqrt{g_{aib}P_a}\sum\limits_{m = 1}^M
%\int_{0}^{+\infty}|h^H_{ib}|_mf_1(|h^H_{ib}|_m)\int_{0}^{+\infty}|h_{ai}|_m\bullet \nonumber\\
%&f_1(|h_{ai}|_m)d(|h_{ai}|_m)d(|h^H_{ib}|_m)+\sqrt{g_{ab}P_a}\int_{0}^{+\infty}|h^H_{ab}|\bullet \nonumber\\
%&f_1(|h^H_{ab}|)d(|h^H_{ab}|)\Big)x+n_b \nonumber\\
&=\left(\sqrt{g_{aib}P_a}M\frac{\pi}{2}\alpha_{ai}\alpha_{ib}
+\sqrt{\frac{g_{ab}P_a\pi}{2}}\alpha_{ab}\right)x+n_b.
\end{align}
\subsection{Performance Loss  of SNR at Bob}
Based on (\ref{y_b_LNQ}), (\ref{y_b_LQE1}), and (\ref{y_b_LQE2}), the SNR expressions of no PL, PL, and APL are given by
\begin{align}
\text{SNR}^{RL}=\frac{\left(\sqrt{g_{aib}P_a}M\frac{\pi}{2}\alpha_{ai}\alpha_{ib}
+\sqrt{\frac{g_{ab}P_a\pi}{2}}\alpha_{ab}\right)^2}{\sigma^2},
\end{align}
\begin{align}
\widehat{\text{SNR}}^{RL}=\frac{\left(\sqrt{g_{aib}P_a}M
\text{sinc}\left(\frac{\pi}{2^k}\right)\frac{\pi}{2}\alpha_{ai}\alpha_{ib}+
\sqrt{\frac{g_{ab}P_a\pi}{2}}\alpha_{ab}\right)^2}{\sigma^2},
\end{align}
and
\begin{align}
&\widetilde{\text{SNR}}^{RL}=\nonumber\\
&\frac{\Big(\sqrt{g_{aib}P_a}M
\big(1-\frac{1}{6}\big(\frac{\pi}{2^k}\big)^2\big)\frac{\pi}{2}\alpha_{ai}\alpha_{ib}+
\sqrt{\frac{g_{ab}P_a\pi}{2}}\alpha_{ab}\Big)^2}{\sigma^2},
\end{align}
respectively.

Then the SNR PL and APL are given as 
\begin{align}
\widehat{L}^{RL}&=\frac{\text{SNR}^{RL}}{\widehat{\text{SNR}}^{RL}}\nonumber\\
&=\frac{\left(\sqrt{g_{aib}}M\frac{\pi}{2}\alpha_{ai}\alpha_{ib}
+\sqrt{\frac{g_{ab}\pi}{2}}\alpha_{ab}\right)^2}
{\left(\sqrt{g_{aib}}M
\text{sinc}\left(\frac{\pi}{2^k}\right)\frac{\pi}{2}\alpha_{ai}\alpha_{ib}+
\sqrt{\frac{g_{ab}\pi}{2}}\alpha_{ab}\right)^2},
\end{align}
and
\begin{align}
\widetilde{L}^{RL}&=\frac{\text{SNR}^{RL}}{\widetilde{\text{SNR}}^{RL}}\nonumber\\
&=\frac{\left(\sqrt{g_{aib}}M\frac{\pi}{2}\alpha_{ai}\alpha_{ib}
+\sqrt{\frac{g_{ab}\pi}{2}}\alpha_{ab}\right)^2}{\left(\sqrt{g_{aib}}M
\left(1-\frac{1}{6}\left(\frac{\pi}{2^k}\right)^2\right)\frac{\pi}{2}\alpha_{ai}\alpha_{ib}+
\sqrt{\frac{g_{ab}\pi}{2}}\alpha_{ab}\right)^2},
\end{align}
respectively.

\subsection{Performance Loss  of Achievable Rate at Bob}
In accordance with (\ref{y_b_LNQ}), (\ref{y_b_LQE1}), and (\ref{y_b_LQE2}), the achievable rates at Bob in the absence of PL,  in the presence of PL and APL are given by
\begin{align}
R^{RL}=\text{log}_2\Bigg(1+\frac{\Big(\sqrt{g_{aib}P_a}M\frac{\pi}{2}\alpha_{ai}\alpha_{ib}
+\sqrt{\frac{g_{ab}P_a\pi}{2}}\alpha_{ab}\Big)^2}{\sigma^2}\Bigg),
\end{align}
\begin{align}
&\widehat{R}^{RL}=\nonumber\\
&\text{log}_2\Bigg(1+\frac{\Big(\sqrt{g_{aib}P_a}M
\text{sinc}\big(\frac{\pi}{2^k}\big)\frac{\pi}{2}\alpha_{ai}\alpha_{ib}+
\sqrt{\frac{g_{ab}P_a\pi}{2}}\alpha_{ab}\Big)^2}{\sigma^2}\Bigg),
\end{align}
and
\begin{align}
&\widetilde{R}^{RL}=\text{log}_2\Bigg(1+\nonumber\\
&\frac{\Big(\sqrt{g_{aib}P_a}M
\Big(1-\frac{1}{6}\left(\frac{\pi}{2^k}\right)^2\Big)\frac{\pi}{2}\alpha_{ai}\alpha_{ib}+
\sqrt{\frac{g_{ab}P_a\pi}{2}}\alpha_{ab}\Big)^2}{\sigma^2}\Bigg),
\end{align}
respectively.

\subsection{BER Performance Loss at Bob}
From (\ref{y_b_LNQ}), (\ref{y_b_LQE1}), and (\ref{y_b_LQE2}), when the QPSK modulation is assumed to be employed, the BERs at Bob with no PL, PL, and APL are given by
\begin{align}
&\text{BER}^{RL}\approx \nonumber\\
&Q\left(\sqrt{\frac{2\Big(\sqrt{g_{aib}P_a}M\frac{\pi}{2}\alpha_{ai}\alpha_{ib}
+\sqrt{\frac{g_{ab}P_a\pi}{2}}\alpha_{ab}\Big)^2}{\sigma^2}}\right),
\end{align}
\begin{align}
&\widehat{\text{BER}}^{RL}\approx \nonumber\\
&Q\left(\sqrt{\frac{2\Big(\sqrt{g_{aib}P_a}M
\text{sinc}\left(\frac{\pi}{2^k}\right)\frac{\pi}{2}\alpha_{ai}\alpha_{ib}+
\sqrt{\frac{g_{ab}P_a\pi}{2}}\alpha_{ab}\Big)^2}{\sigma^2}}\right),
\end{align}
and
\begin{align}
&\widetilde{\text{BER}}^{RL}\approx \nonumber\\
&Q\Bigg(\sqrt{\frac{2 P_a\big(\sqrt{g_{aib}}M
\big(1-\frac{1}{6}\big(\frac{\pi}{2^k}\big)^2\big)\frac{\pi}{2}\alpha_{ai}\alpha_{ib}+
\sqrt{\frac{g_{ab}\pi}{2}}\alpha_{ab}\big)^2}{\sigma^2}}\Bigg),
\end{align}
respectively. It is noted that the above derived results may be extended to the scenarios of high-order digital modulations like M-ary phase shift keying (MPSK), M-ary quadrature amplitude modulation (MQAM).

\section{Simulation Results and Discussions}
In this section, simulation results are presented to evaluate the effect of phase mismatching caused by IRS with discrete phase shifters from three different aspects: SNR, AR, and BER. The path loss at the distance $\bar{d}$ is modeled as $g(\bar{d})=\text{PL}_0-10\gamma\text{log}_{10}\frac{\bar{d}}{d_0}$, where $\text{PL}_0=-30$dB represents the path loss reference distance $d_0=1$m, and $\gamma$ is the path loss exponent. The path loss exponents of Alice$\rightarrow$IRS, IRS$\rightarrow$Bob, and Alice$\rightarrow$Bob channels are respectively chosen as 2, 2, and 2 in the LoS channels, and the one are respectively set to be 2.5, 2.5, and 3.5 in the Rayleigh channels. The default system parameters are chosen as follows: $M=128$, $d=\lambda/2$, $\theta_{ab}=\pi/2$, $\theta_{ai}=\pi/4$, $d_{ab}=100$m, $d_{ai}=30$m, $P_a=30$dBm.
$\alpha_{ai}=\alpha_{ib}=\alpha_{ab}=0.5$.

%(QPSK)
%$N=16$, $K=128$, $N_b=16$, $d=\lambda/2$, $\beta=0.5$, $\alpha=0.8$, $c=2$, $\theta_{b}=60\textordmasculine$, $\theta_{e}=120\textordmasculine$, $\theta_{be}=45\textordmasculine$, $d_{ab}=d_{ae}=d_{be}=500 m$, $P_a=P_b=70 dBm$, and modulation scheme being QPSK.
%\subsection{Perfrmance loss }
%\subsection{Proposed RP}
%\subsection{Secure performance analysis}

\begin{figure}[htbp]
\centering

\subfigure [LoS channels]{
\begin{minipage}[]{0.52\textwidth}
\centering
  \includegraphics[width=1\textwidth]{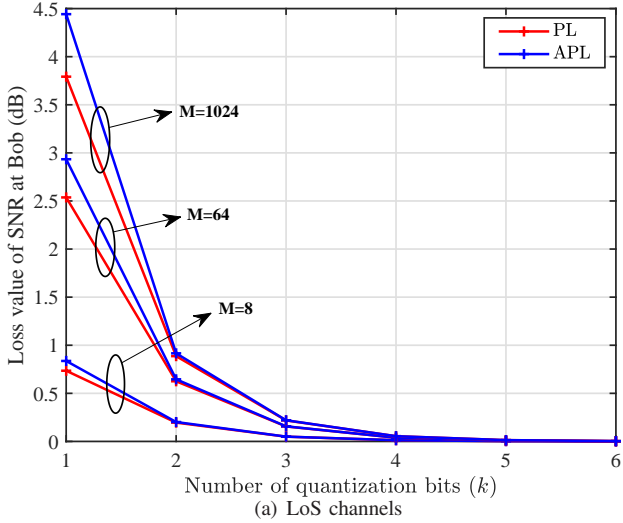}\\
\end{minipage}
}

\subfigure[Rayleigh channels]{
\begin{minipage}[]{0.52\textwidth}
\centering
  \includegraphics[width=1\textwidth]{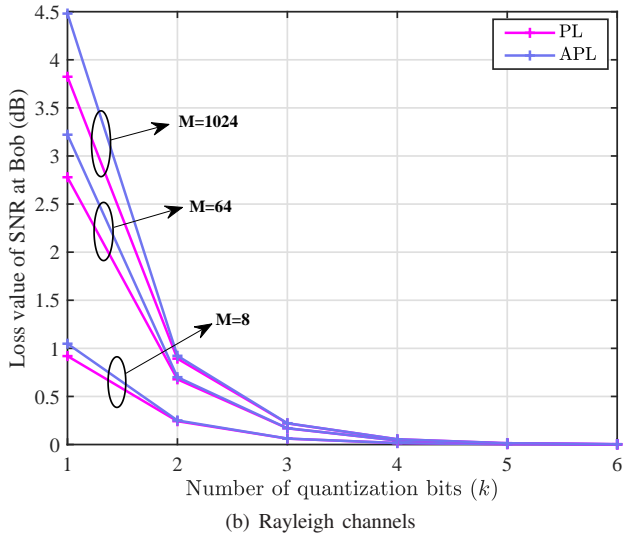}\\
\end{minipage}
}

\caption{Curves of loss of SNR versus the number $k$ of quantization bits.}\label{SNR}
\end{figure}

Figs.~\ref{SNR} (a)  and (b) plot the curves of SNR performance loss versus the number $k$ of quantization bits ranging from 1 to 6 in LoS and Rayleigh channels, respectively, where three different IRS element numbers $M$ are chosen: 8, 64 and 1024. It can be seen from the  two subfigures that regardless of the case of performance loss (PL) or APL, the SNR performance loss in the LoS channels and Rayleigh channels decreases as the number of quantization bits $k$ increases, while it increases with $M$ increases. In addition, when $k$ is larger than or equal to 3, the SNR performance loss is less than 0.23dB even when the number of IRS phase shift elements $M$ tends to large scale (e.g., $M=1024$). This means that 3 bits is sufficient to achieve a trivial performance loss.

%\begin{figure*}
% \setlength{\abovecaptionskip}{-5pt}
% \setlength{\belowcaptionskip}{-10pt}
% \centering
% \begin{minipage}[t]{0.33\linewidth}
%  \centering
%  \includegraphics[width=2.5in]{SINR.eps}
%  \caption{Curves of loss of SINR versus $q$ for different values of $\eta$.}
% \end{minipage}%
% \begin{minipage}[t]{0.33\linewidth}
%  \centering
%  \includegraphics[width=2.5in]{figure32.eps}
%  \caption{Curves of SR versus $q$ for different values of $\eta$.}
% \end{minipage}
% \begin{minipage}[t]{0.33\linewidth}
%  \centering
%  \includegraphics[width=2.5in]{figure42.eps}
%  \caption{Curves of BER versus degree for different values of $\eta$ and $q$.}
% \end{minipage}
%\end{figure*}
%\begin{figure}[h]
%\centering
%\includegraphics[width=0.423\textwidth]{SINR.eps}\\
%%\includegraphics[width=0.45\textwidth]{figure22.eps}\\
%\caption{\textcolor{blue}{Curves of loss of SINR versus $q$ for different values of $\eta$.}}%\label{system-model.eps}
%\end{figure}
%\begin{figure}[h]
%\centering
%\includegraphics[width=0.445\textwidth]{figure32.eps}\\
%\caption{\textcolor{blue}{Curves of SR versus $q$ for different values of $\eta$.}}%\label{system-model.eps}
%\end{figure}
%\begin{figure}[h]
%\centering
%\includegraphics[width=0.443\textwidth]{figure42.eps}\\
%\caption{\textcolor{blue}{Curves of BER versus degree for different values of $\eta$ and $q$.}}%\label{system-model.eps}
%\end{figure}

\begin{figure}[htbp]
\centering

\subfigure [LoS channels]{
\begin{minipage}[]{0.52\textwidth}
\centering
  \includegraphics[width=1\textwidth]{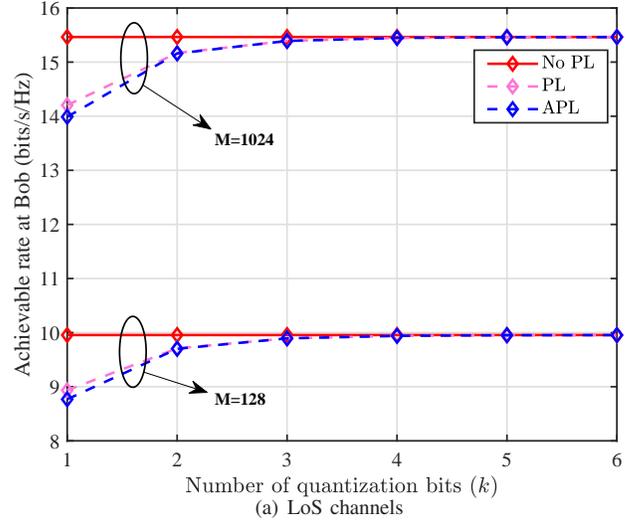}\\
\end{minipage}
}

\subfigure[Rayleigh channels]{
\begin{minipage}[]{0.52\textwidth}
\centering
  \includegraphics[width=1\textwidth]{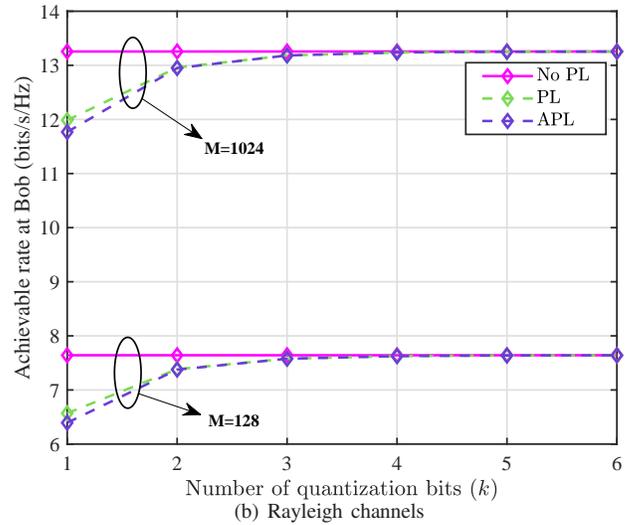}\\
\end{minipage}
}

 \caption{Curves of AR versus the number $k$ of quantization bits.}\label{AR}
\end{figure}

Figs.~\ref{AR} (a)  and (b) show the  curves of AR versus the number $k$ of quantization bits ranging from 1 to 6 in LoS and Rayleigh channels, respectively, where SNR is equal to 15dB. From Fig.~\ref{AR}, it is seen that
%the PL of SR at Bob increases as $\eta$ increases,
the AR performance loss at Bob decreases as $k$ increases, and increases as $M$ increases. Additionally, the AR increases as the number of IRS phase shift elements $M$ increases. Compared with the case of no PL,
3 quantization bits achieves a AR performance loss less than 0.08 bits/s/Hz in the cases of PL and APL regardless of the number of IRS phase shift elements. When the number of quantization bits is larger than or equal to 2, the simple approximate PL expression coincides with the true performance loss.

%\begin{figure}[htbp]
%\centering
%\includegraphics[width=0.5\textwidth]{AR_0.eps}\\
%%\includegraphics[width=0.45\textwidth]{figure22.eps}\\
%\caption{Curves of AR versus the number $k$ of quantization bits (SNR=0dB).}\label{AR1}
%\end{figure}
%\begin{figure}[htbp]
%\centering
%\includegraphics[width=0.5\textwidth]{AR_15.eps}\\
%%\includegraphics[width=0.45\textwidth]{figure22.eps}\\
%\caption{Curves of AR versus the number $k$ of quantization bits (SNR=15dB).}\label{AR2}
%\end{figure}

\begin{figure}[htbp]
\centering
\includegraphics[width=0.52\textwidth]{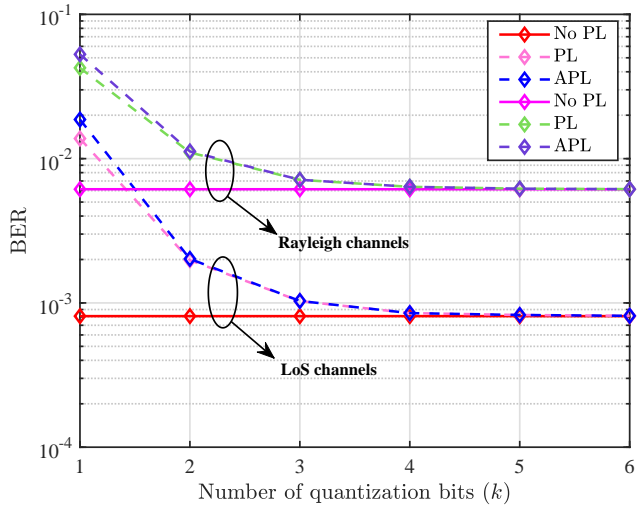}\\
\caption{Curves of BER versus the number $k$ of quantization bits.}\label{BER}
\end{figure}

Fig.~\ref{BER} illustrates the curves of BER versus the number $k$ of quantization bits from 1 to 6, where SNR is equal to $-5$dB. From Fig.~\ref{BER}, it can be seen that with increasing the number $k$ of quantization bits, the BER performances of PL and APL rapidly approach that no PL. When $k$ reaches up to 3, the BER performances of PL and APL are almost identical to that of no PL, which means that it is feasible in practice to use discrete phase shifters with $k=3$ to achieve a trivial performance loss. This dramatically reduces the circuit cost and the required CSI amount fed back from BS or user.
%\begin{figure}[htbp]
%\centering
%\includegraphics[width=0.5\textwidth]{BER.eps}\\
%%\includegraphics[width=0.45\textwidth]{figure22.eps}\\
%\caption{Curves of SNR versus the number $k$ of quantization bits ($d_{ab}=200$m).}\label{d2}
%\end{figure}
\section{Conclusion}
In this paper, the performance of IRS with discrete phase shifters of wireless network has been investigated. To  make an analysis of  the
performance loss caused by IRS with phase quantization error, we considered two scenarios: LoS and Rayleigh channels. The closed-form expressions of SNR performance loss, AR, and BER with PL were derived using the law of large numbers and some mathematic approximation techniques. With the help of the Taylor series expansion, the simple approximate performance loss  expressions of IRS with approximate quantization error were also provided. Simulation results showed that when the number of quantization bits is larger than or equal to 3, the performance losses of SNR and AR are less than 0.23dB and 0.08bits/s/Hz, respectively, and the corresponding degradation on BER is negligible. The simple approximate expression approaches the true performance loss when the number of quantization bits is larger than or equal to 2.
% In other words, adjusting the phase matrix of IRS requires about 3$M$-bit CSI, which will significantly reduce the required CSI amount from BS and users and circuit cost.
%The  performance loss analysis can serve to guide how to choose  a cost-effective IRS with discrete-phase shifters. 
%The derived performance loss expressions will be applied to decide the proper number of quantization bits in a practical IRS-aided wireless network in the coming future so as to reduce the circuit cost. 
%In summary, 3-bit is sufficient for IRS-aided wireless networks.
%\section{Numerical results}

%\section{Conclusions}

\ifCLASSOPTIONcaptionsoff
  \newpage
\fi

\bibliographystyle{IEEEtran}
\bibliography{IEEEfull,reference}

\begin{thebibliography}{10}
\providecommand{\url}[1]{#1}
\csname url@samestyle\endcsname
\providecommand{\newblock}{\relax}
\providecommand{\bibinfo}[2]{#2}
\providecommand{\BIBentrySTDinterwordspacing}{\spaceskip=0pt\relax}
\providecommand{\BIBentryALTinterwordstretchfactor}{4}
\providecommand{\BIBentryALTinterwordspacing}{\spaceskip=\fontdimen2\font plus
\BIBentryALTinterwordstretchfactor\fontdimen3\font minus
  \fontdimen4\font\relax}
\providecommand{\BIBforeignlanguage}[2]{{%
\expandafter\ifx\csname l@#1\endcsname\relax
\typeout{** WARNING: IEEEtran.bst: No hyphenation pattern has been}%
\typeout{** loaded for the language `#1'. Using the pattern for}%
\typeout{** the default language instead.}%
\else
\language=\csname l@#1\endcsname
\fi
#2}}
\providecommand{\BIBdecl}{\relax}
\BIBdecl

\bibitem{Nguyen2022Achievable}
V.~Nguyen, T.~P. Truong, T.~M.~T. Nguyen, W.~Noh, and S.~Cho, ``Achievable rate
  analysis of two-hop interference channel with coordinated {IRS} relay,''
  \emph{IEEE Trans. Wirel. Commun.}, pp. 1--1, 2022.

\bibitem{Wu2020Towards}
Q.~Wu and R.~Zhang, ``Towards smart and reconfigurable environment: Intelligent
  reflecting surface aided wireless network,'' \emph{IEEE Commun Mag}, vol.~58,
  no.~1, pp. 106--112, Jan. 2020.

\bibitem{Tang2020MIMO}
W.~Tang, J.~Dai, M.~Chen, K.~Wong, X.~Li, X.~Zhao, S.~Jin, Q.~Cheng, and
  T.~Cui, ``{MIMO} transmission through reconfigurable intelligent surface:
  system design, analysis, and implementation,'' \emph{IEEE J. Sel. Areas
  Commun.}, vol.~38, no.~11, pp. 2683--2699, Nov. 2020.

\bibitem{Gao2021Reflection}
Y.~Gao, C.~Yong, Z.~Xiong, J.~Zhao, Y.~Xiao, and D.~Niyato, ``Reflection
  resource management for intelligent reflecting surface aided wireless
  networks,'' \emph{IEEE Trans. Commun.}, vol.~69, no.~10, pp. 6971--6986, Oct.
  2021.

\bibitem{Huang2019Reconfigurable}
C.~Huang, A.~Zappone, G.~C. Alexandropoulos, M.~Debbah, and C.~Yuen,
  ``Reconfigurable intelligent surfaces for energy efficiency in wireless
  communication,'' \emph{IEEE Trans. Wirel. Commun.}, vol.~18, no.~8, pp.
  4157--4170, Aug. 2019.

\bibitem{Guan2020Intelligent}
X.~Guan, Q.~Wu, and R.~Zhang, ``Intelligent reflecting surfaces assisted
  secrecy commuincation: Is artificial noise helpful or not?'' \emph{IEEE
  Wireless Commun. Lett.}, vol.~9, no.~6, pp. 778--782, Jun. 2020.

\bibitem{Dong2022Beamforming}
R.~Dong, F.~Shu, R.~Chen, Y.~Wu, C.~Pan, and J.~Wang, ``Beamforming and power
  allocation for double-{RIS}-aided two-way directional modulation network,''
  [online] Available: https://arxiv.org/abs/2201.09063.

\bibitem{Hong2020Artificial}
S.~Hong, C.~Pan, H.~Ren, K.~Wang, and A.~Nallanathan, ``Artificial-noise-aided
  secure {MIMO} wireless communications via intelligent reflecting surface,''
  \emph{IEEE Trans. Commun.}, vol.~68, no.~12, pp. 7851--7866, Dec. 2020.

\bibitem{Wang2022Beamforming}
X.~Wang, F.~Shu, W.~Shi, X.~Liang, R.~Dong, J.~Li, and J.~Wang, ``Beamforming
  design for {IRS}-aided decode-and-forward relay wireless network,''
  \emph{IEEE Trans. Green Commun. Netw.}, vol.~6, no.~1, pp. 198--207, Mar.
  2022.

\bibitem{Zhu2022Intelligent}
Q.~Zhu, Y.~Gao, Y.~Xiao, and S.~Mumtaz, ``Intelligent reflecting surface aided
  wireless networks: Dynamic user access and system sum-rate maximization,''
  \emph{IEEE Trans. Commun.}, pp. 1--1, 2022.

\bibitem{Ahmed2022Reconfigurable}
A.-H. Ahmed, M.~Samir, M.~Elhattab, C.~Assi, and S.~Sharafeddine,
  ``Reconfigurable intelligent surface enabled vehicular communication: Joint
  user scheduling and passive beamforming,'' \emph{IEEE Trans. Veh. Technol.},
  vol.~71, no.~3, pp. 2333--2345, Mar. 2022.

\bibitem{Mei2021Performance}
W.~Mei and R.~Zhang, ``Performance analysis and user association optimization
  for wireless network aided by multiple intelligent reflecting surface,''
  \emph{IEEE Trans. Commun.}, vol.~69, no.~9, pp. 6296--6312, Sep. 2021.

\bibitem{Yang2020Secrecy}
L.~Yang, J.~Yang, W.~Xie, M.~O. Hasna, T.~Tsiftsis, and M.~D. Renzo, ``Secrecy
  performance analysis of {RIS}-aided wireless communication systems,''
  \emph{IEEE Trans. Veh. Technol.}, vol.~69, no.~10, pp. 12\,296--12\,300,
  2020.

\bibitem{Fang2021Joint}
S.~Fang, G.~Chen, and Y.~Li, ``Joint optimization for secure intelligent
  reflecting surface assisted {UAV} networks,'' \emph{IEEE Commun. Lett.},
  vol.~10, no.~2, pp. 276--280, May. 2021.

\bibitem{ShuEnhanced2021}
F.~Shu, Y.~Teng, J.~Li, M.~Huang, W.~Shi, J.~Li, Y.~Wu, and J.~Wang, ``Enhanced
  secrecy rate maximization for directional modulation networks via {IRS},''
  \emph{IEEE Trans. Commun.}, vol.~69, no.~12, pp. 8388--8401, Dec. 2021.

\bibitem{Di2020Hybrid}
B.~Di, H.~Zhang, L.~Song, Y.~Li, Z.~Han, and H.~V. Poor, ``Hybrid beamforming
  for reconfigurable intelligent surface based multi-user communications:
  achievable rates with limited discrete phase shifts,'' \emph{IEEE J. Sel.
  Areas Commun.}, vol.~38, no.~8, pp. 1809--1822, Aug. 2020.

\bibitem{Choi2021Multiple}
J.~Choi, G.~Kwon, and H.~Park, ``Multiple intelligent reflecting surfaces for
  capacity maximization in {LOS} {MIMO} systems,'' \emph{IEEE Wireless Commun.
  Lett.}, vol.~10, no.~8, pp. 1727--1731, Aug. 2021.

\bibitem{Rehman2021Joint}
H.~U. Rehman, F.~Bellili, A.~Mezghani, and E.~Hossain, ``Joint active and
  passive beamforming design for {IRS}-assisted multi-user {MIMO} systems: a
  {VAMP}-based approach,'' \emph{IEEE Trans. Commun.}, vol.~69, no.~10, pp.
  6734--6749, Oct. 2021.

\bibitem{Han2022Double}
Y.~Han, S.~Zhang, L.~Duan, and R.~Zhang, ``Double-{IRS} aided {MIMO}
  communication under {LoS} channel: capacity maximization and scaling,''
  \emph{IEEE Trans. Wirel. Commun.}, pp. 1--1, 2022.

\bibitem{Wang2020Intelligent}
H.-M. Wang, J.~Bai, and D.~Limeng, ``Intelligent reflecting surfaces assisted
  secure transmission without eavesdropper's {CSI},'' \emph{IEEE Signal Process
  Lett.}, vol.~27, pp. 1300--1304, 2020.

\bibitem{Wu2019Beamforming}
Q.~Wu and R.~Zhang, ``Beamforming optimization for intelligent reflecting
  surface with discrete phase shifts,'' \emph{Proc.IEEE ICASSP}, pp.
  7830--7833, May. 2019.

\bibitem{Shi2021Secure}
W.~Shi, J.~Li, G.~Xia, Y.~Wang, X.~Zhou, Y.~Zhang, and F.~Shu, ``Secure
  multigroup multicast communication systems via intelligent reflecting
  surface,'' \emph{China Communications}, vol.~18, no.~3, pp. 39--51, Mar.
  2021.

\bibitem{Zhang2020Sum}
Y.~Zhang, C.~Zhong, Z.~Zhang, and W.~Lu, ``Sum rate optimization for two way
  communications with intelligent reflecting surface,'' \emph{IEEE Commun.
  Lett.}, vol.~24, no.~5, pp. 1090--1094, May. 2020.

\bibitem{Shen2020Beamforming}
H.~Shen, T.~Ding, W.~Xu, and C.~Zhao, ``Beamforming design with fast
  convergence for {IRS}-aided full-duplex communication,'' \emph{IEEE Commun.
  Lett.}, vol.~24, no.~12, pp. 2849--2853, Dec. 2020.

\bibitem{Pan2020Multicell}
C.~Pan, H.~Ren, K.~Wang, W.~Xu, M.~Elkashlan, A.~Nallanathan, and L.~Hanzo,
  ``Multicell {MIMO} communications relying on intelligent reflecting
  surfaces,'' \emph{IEEE Trans. wirel. Commun.}, vol.~19, no.~8, pp.
  5218--5233, Aug. 2020.

\bibitem{Li2019Performance}
J.~Li, L.~Xu, P.~Lu, T.~Liu, Z.~Zhang, J.~Hu, F.~Shu, and W.~J., ``Performance
  analysis of directional modulation with finite{-}quantized {RF} phase
  shifters in analog beamforming structure,'' \emph{IEEE Access}, vol.~7, pp.
  97\,457--97\,465, July. 2019.

\bibitem{Wei2021Secure}
Z.~Wei, C.~Masouros, and F.~Liu, ``Secure directional modulation with few-bit
  phase shifters: optimal and iterative-closed-form designs,'' \emph{IEEE
  Trans. Commun.}, vol.~69, no.~1, pp. 486--500, Jan. 2021.

\bibitem{Dong2022Performance}
R.~Dong, B.~Shi, X.~Zhan, F.~Shu, and J.~Wang, ``Performance analysis of
  massive hybrid directional modulation with mixed phase shifters,'' \emph{IEEE
  Trans. Veh. Technol.}, pp. 1--1, 2022.

\bibitem{You2020Channel}
C.~You, B.~Zheng, and R.~Zhang, ``Channel estimation and passive beamforming
  for intelligent reflecting surface: Discrete phase shift and progressive
  refinement,'' \emph{IEEE J. Sel. Areas Commun.}, vol.~38, no.~11, pp.
  2604--2620, Nov. 2020.

\bibitem{Wasserman2004All}
L.~Wasserman, ``All of statistics: A concise course in statistical inference,''
  \emph{New York, NY, USA: Springer}, 2004.

\bibitem{Moon1999Mathematical}
T.~K. Moon and W.~C. Stirling, ``Mathematical methods and algorithms for signal
  processing,'' \emph{USA: Marsha Horron}, 1999.

\end{thebibliography}
\end{document}